\documentclass[onecolumn,doublespacing]{elsart}

\usepackage[dvips]{graphicx}
\usepackage{amsbsy}
\usepackage{natbib}
\usepackage{epsfig}
\usepackage{mathenv}
\usepackage{array}
\usepackage{color}
\usepackage{float}
\usepackage{longtable}
\usepackage{lscape}
\usepackage[counterclockwise]{rotating}
\usepackage{indentfirst}

\begin{document}
\hyphenpenalty 10000
\bibliographystyle{pepi}

\begin{frontmatter}

\title{Constraints on grain size and stable iron phases in the uppermost Inner Core
from multiple scattering modeling of seismic velocity and attenuation}

%
%

\author[LGIT,DTP]{Marie Calvet\corauthref{cor}}
\corauth[cor]{Corresponding author}
\ead{calvet@dtp.obs-mip.fr}
\author[CEREGE]{Ludovic Margerin}
\ead{margerin@cerege.fr}
\address[LGIT]{Laboratoire de G\'eophysique Interne et Tectonophysique, BP 53, 38041 Grenoble cedex 9, FRANCE}
\address[DTP]{Now at Laboratoire de Dynamique Terrestre et Plan\'etaire, Observatoire Midi-Pyr\'en\'ees, 14 avenue Edouard Belin, 31400 Toulouse, FRANCE}
\address[CEREGE]{Centre Europ\'een de Recherche et d'Enseignement des G\'eosciences de l'Environnement, BP 80, 13545 Aix-en-Provence, FRANCE}

%
%
\begin{abstract}
We propose to model the uppermost
inner core as an aggregate of randomly oriented anisotropic ``patches''. A patch is defined
as an assemblage of a possibly large number of crystals with identically oriented crystallographic axes. 
 This simple model accounts for the observed velocity isotropy of short period body waves,
  and offers a reasonable physical interpretation for the 
scatterers detected at the top of the inner core.  From rigorous multiple scattering modeling of seismic wave propagation
through the aggregate, we obtain  strong constraints on both  the size and the 
elastic constants of iron patches. In a first step, we study the phase velocity
and scattering attenuation  of aggregates composed of hexagonal and cubic crystals, whose elastic constants have been
 published in the mineral physics literature.
The predicted attenuations for {\it P} waves vary over two orders of magnitude.
Our calculations demonstrate  that scattering attenuation is extremely sensitive to the anisotropic properties of single crystals and offers
an attractive way to discriminate among iron models with e.g. identical Voigt average speeds.
When  anisotropy of elastic constants is pronounced, we find that the {\it S} wavespeed in the aggregate can be as much
as 15\% lower than the Voigt average shear velocity of  a single crystal.
In a second step, we perform a systematic search for iron models compatible with measured seismic velocities
and attenuations. An iron model is characterized by its symmetry (cubic or hexagonal), elastic constants, and patch size. 
Independent of the crystal symmetry, we infer a most likely size of patch of the order of 400~m. 
  Recent {\it bcc} iron models from the literature are in very good  agreement with the most probable elastic constants
  of cubic crystals found in our inversion. Our study (1) 
 suggests that the presence of  melt may not be required to explain the 
 low shear wavespeeds in the inner core and (2) supports the recent experimental results on the stability of cubic iron in
the inner core,  at least in its upper part.

\end{abstract}
\end{frontmatter}
\section{Introduction}
The gross seismic features of the inner core are nowadays known with some confidence.
The uppermost inner core, where crystallization
of iron occurs, is a peculiar region with isotropic seismic velocities, strong attenuation and where the presence of scatterers has been detected.
Mineral physicists have concentrated a lot of efforts on experimental and theoretical investigations
 of  elastic properties of iron at inner core condition. In spite of remarkable advances, fundamental
 questions pertaining to  the symmetry class and the anisotropy parameters of iron are still
 actively debated and no clear consensus has emerged yet. In this study, we use the velocity
 and attenuation properties of seismic waves inferred from previous studies to propose a simple
 physical model of the uppermost inner core.  Using a rigorous multiple scattering approach based
 on the Dyson equation for elastic waves in random media, we analyze critically various iron
 models from the mineral physics  literature and give constraints on the possible stable iron
 phases in the inner core. In what follows, the necessary seismological and mineralogical data are reviewed and our
 approach is described.  
\subsection{Seismic observations and mineral physics }

Traveltime analyses of seismic body waves  have revealed that the Earth's
 inner core is anisotropic with about 1-3\% 
 velocity anisotropy and a fast direction of propagation parallel to Earth's rotation axis.
Depth variations have been observed with an isotropic layer overlying deep anisotropy
\cite[]{Shearer94,Sudziew95,creager99,Garsou00,Wen02} and a change of velocity 
anisotropy near the centre
 of the inner core \cite[]{Ishii02,Begtramp03,calvet1b}.
The inner core is heterogeneous at different scales with hemispherical velocity variations
 \cite[]{Niuwen01,Garcia02,Cao04}, continent scale \cite[]{strouj04} and short 
wavelength features (20-200~km) \cite[]{Breger99}.
There are strong evidences for the concentration of very small scale heterogeneities at the top of the inner core.
The {\it PKIKP} wave that bottoms in the uppermost inner core is strongly attenuated compared to outer core phases. In addition,
its frequency spectrum is depleted in high frequencies. Various studies proposed a range of values for {\it P}-wave quality factor between 100
and 400 in the uppermost 200~km with possible lateral variations \cite[e.g.][]{Souriau95,Wen02,Cao04,Yu06b}. Viscoelastic and scattering attenuation
are candidates to explain this observation \cite[]{Licorm02,cormli02} but the latter mechanism seems more consistent
with the strong coda of the reflected {\it P} wave generated at the inner core boundary
 \cite[]{VidaleNat,Poup04,Koper04,Krasno05,Leyton07b}.  Thus, the prominent features of the uppermost inner core are its
 isotropy and strong scattering properties. An important open question is to relate these macroscopic 
 properties to the microscopic anisotropy of the iron alloy. 

Iron is thought to be the main constituent of Earth's inner core. However, the bulk properties and crystalline
 structure of Fe at such extreme physical conditions remain uncertain.
The hexagonal-close-packed ({\it hcp}) phase of iron
has been proposed as a likely stable phase in the inner core by several authors
 \citep[e.g.][]{Stix95,Yoo95, Ma04} but there is no consensus on its anisotropy \cite[]{Mao98,Stein01,Antona06}.
  Because the uncertainty on the temperature in the inner core
 is very large \cite[]{Wil87,Boeh93,Saxena94,Ma04}, other structures such as body-centered cubic
 \cite[]{Poishank93,Saxena98}, double hexagonal
 closed packed \cite[]{Saxdubrov95} and orthorhombic \cite[]{Anderduba97,Andrault00} cannot  be excluded.
 The problem is made even more complex by the nature and quantity of light elements present
 in the inner core \cite[]{Poirier94}. 
Nickel or silicon may stabilise a
body-centered-cubic ({\it bcc}) phase at inner core conditions \cite[]{Lin02b,Voca03,Voca07,Dubro07}.
A long standing issue in mineral physics is the discrepancy between the Voigt average
velocity of iron crystals and the observed seismic velocity. 
Even the most recent models such
as proposed by \citet{Voca07} and \citet{Belono07}, that incorporate the effect of temperature, have Voigt average shear 
velocity between 4.0 to  4.4 km.s$^{-1}$.
This is still significantly higher than the adopted 3.5 km.s$^{-1}$ velocity in seismological models.

\subsection{Approach of this study}
Two distinct classes of model have been proposed to explain the bulk seismic properties 
of the solid core:  (1)  Some authors have invoked the presence of  elliptic liquid inclusions
which could mimic the observed anisotropy and attenuation of seismic waves through the inner core \cite[]{Singh00};
(2) \citet{Bergman97} proposed a solidification texturing model which results in depth dependent anisotropic
   properties in qualitative agreement with body wave traveltime measurements. 
In this work, we focus on the specific scattering properties of the uppermost inner core and its implication
for the attenuation and velocity dispersion of seismic waves.
Previous seismological studies of scattering  have thus far only considered the effects of
elastically isotropic heterogeneities  with possibly  anisotropic (anisomeric)  spatial distributions \cite[]{Cormier98,VidaleNat,cormli02,Cormier07}.
These seismological models of heterogeneity are not clearly connected with the mineralogical and geodynamical
models of the inner core. One may for instance ask: what is the physical meaning of an isotropic velocity perturbation in an intrinsically
anisotropic material? 
 
\begin{figure}
\center
\includegraphics[width=14.0cm]{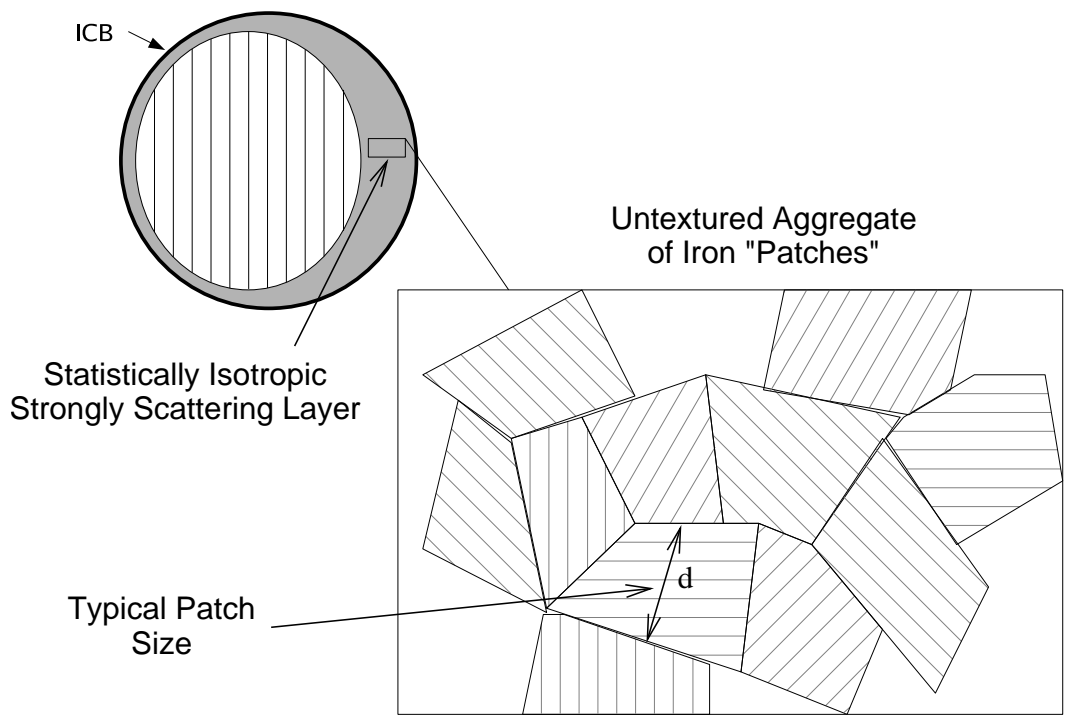}
\caption{Modelling of the superficial layer at the top of the solid inner core as an untextured aggregate of iron
 ``patches'', with typical size $d$. ICB: Inner Core Boundary.}
\label{agregat}
\end{figure}

We propose a simple model of  solidification texturing of the uppermost 
inner core, consistent with seismological and mineralogical data. We limit our considerations to a few simple quantities 
that can be inferred from seismic 
observations: {\it P} and {\it S} wave velocities and {\it P}-wave attenuation. 
We model the superficial part of the solid core as an
 untextured aggregate of iron ``patches'' \cite[]{Krasno05} as depicted in Figure \ref{agregat}. Each patch is characterized
 by the anisotropic  properties of individual iron crystals and the orientation of crystallographic axes
varies randomly from one patch to another. One patch is not necessarily made
of a single crystal. One can for example imagine that it contains a large number
of dendrites with strongly correlated orientation of crystallographic axes.  
Such a texture is shown in Figure \ref{agregat} and is similar to what is observed
in laboratory experiments of crystallization of  ice or {\it hcp} iron
  \cite[]{Worster97,Bergman97,Bergman03}.
The seismic properties of the aggregate depend on the patch size and  the iron crystal
 properties (symmetry, anisotropy). As a consequence  of our assumption of random patch orientation, the
 aggregate is macroscopically isotropic.
But, because of crystal anisotropy, seismic wave velocities vary from one patch to an other which makes
 the aggregate  microscopically inhomogeneous. A seismic pulse travelling through such a medium 
will be prone to amplitude attenuation and velocity dispersion.
We  show how multiple scattering theory can be used to relate the intrinsic elastic properties of iron crystals
to the macroscopic seismic properties of the aggregate. An important conclusion of our theory 
is that the Voigt average velocity of a single crystal is a poor approximation of the seismic velocity of the aggregate.  
\section{Multiple scattering theory}
To describe statistically the aggregate, we introduce the following decomposition
of the elastic tensor: 
\begin{eqnarray}
C_{ijkl}(\mathbf{x}) & = & C^0_{ijkl}\;+\;\delta C_{ijkl}(\mathbf{x})\\
C^0_{ijkl} & = &\langle C_{ijkl}(\mathbf{x})\rangle ,
\end{eqnarray}
where
\begin{eqnarray}
\langle \delta C_{ijkl}(\mathbf{x}) \rangle & = & 0 .
\end{eqnarray}
Angular brackets represent the ensemble average over all possible orientation of a single crystal,
also known as the Voigt average. The fluctuations $ \delta C_{ijkl}(\mathbf{x})$
are considered to be single realizations drawn from an ensemble of random fields, having zero mean.
In such an aggregate, the spatial variations of the elastic constants
 are only caused by different orientations of the patches.
As usual the material heterogeneity   is described by the  correlation of the fluctuations, which in our case
is a tensor field \cite[]{Stankekino}:
\begin{eqnarray}
\langle \delta C_{ijkl}(\mathbf{x})\delta C_{\alpha \beta \gamma \delta}(\mathbf{x'}) \rangle \; = \; \Xi ^{\alpha \beta \gamma \delta} _{ijkl}\eta(|\mathbf{x}-\mathbf{x'}|) ,   \label{correl}
\end{eqnarray}
where $\Xi$ denotes the eigth rank covariance tensor of the  elastic
 moduli and depends on the symmetry of the crystal \cite[]{Hirse88}. The function $\eta(x)$ which equals 1 at $x = 0$ 
  and tends to 0  as $x \to \infty$ is the spatial correlation function. It gives the probability that
 two points separated a distance $x$  from one another are located within the same patch \cite[]{Stanke86}.
 Equation (\ref{correl})  reflects two assumptions regarding
the statistics of the aggregate: first, there are no orientation correlations between different patches 
(i.e. there is no macroscopic anisotropy of the aggregate); second,
the aggregate is statistically homogeneous.
 We have chosen a spatial correlation function of the form:
\begin{eqnarray}
\eta(r)\;=\; e^{-r/a}. \label{correlf}
\end{eqnarray}
The correlation function (\ref{correlf}) implies that the patches (or grains) are convex and equi-axed, i.e., not elongated in some particular
 direction. Such a texture would produce anisotropic velocity and attenuation as shown by  \citet{Margerin06}. In spite of the
 limitations mentioned above, an exponential function describes the variable shapes and linear dimensions of grains in a polycrystalline
  material reasonably well \citep{Stanke86}.  In this case, the effective average dimension of the patches is $ d = 2\,a$.

The inhomogeneity of the aggregate is related to the anisotropy of the patches.
For example, if iron is weakly anisotropic, the elastic properties of the medium have only weak variations
 from patch to patch. The degree of inhomogeneity can be expressed in
terms of the effective elastic constants as \cite[]{Stankekino}:
\begin{eqnarray}
\epsilon ^2 & \simeq & \frac{1}{4} \frac{\langle( C_{IJ}-C^0_{IJ} )^2\rangle}{(C^0_{IJ})^2} . \label{voigt}
\end{eqnarray}
To represent the fourth-order stiffness tensor, we have introduced in equation (\ref{voigt}) the Voigt matrix $C_{IJ}$ 
whose indices $I$ and $J$ vary from 1 to 6 with the following correspondence rule between tensor and matrix indices: 
$ (11)\;\rightarrow 1,\;(22)\;\rightarrow \;2,\;(33)\;\rightarrow\;3,\;(23)\;\rightarrow\;4,\;(13)\;\rightarrow\;5,\;(12)\;\rightarrow \;6$.
The degree of inhomogeneity is defined with $I=J=3$ for {\it P} waves, and with $I=J=4$ for {\it S} waves.

To calculate the seismic response of the aggregate, we have used a formalism based on the
Dyson equation \cite[]{Rytov89} for the ensemble averaged Green function. The theory takes into account all the physics of the problem:
 arbitrary anisotropy of iron, mode conversions between {\it P} and {\it S} waves, and multiple scattering \cite[]{Weaver90}.
This formalism yields effective phase velocities and spatial decay rates of coherent {\it P} and {\it S} waves
propagating through the aggregate.
At a given frequency $\omega$, the seismic coherent field $\langle \mathbf{G}\rangle$
  can be decomposed into  longitudinal ($P$) and transverse ($S$) contributions \cite[]{Weaver90}:
\begin{eqnarray}
\langle \mathbf{G}(\omega , \mathbf{p})\rangle = g^P(p)\mathbf{\hat{p}}\mathbf{\hat{p}}+g^S(p) (\mathbf{I}-\mathbf{\hat{p}}\mathbf{\hat{p}}) ,
\label{dyson}
\end{eqnarray}
where  $g^P(p)$ and $g^S(p)$ are defined as:
\begin{eqnarray}
g^P(p)&=&\frac{1}{\omega^2 -\left(p {V_0^P}\right)^2+\sigma^P(p)}\\
g^S(p)&=&\frac{1}{\omega^2 -\left(p {V_0^S}\right)^2+\sigma^S(p)}.
\end{eqnarray}
In equation (\ref{dyson}), $\mathbf{I}$ is the identity tensor, $\mathbf{\hat{p}}$ is a unit vector in the direction of the wavevector
$\mathbf{p}$; ${V_0^P}$ 
 ($V_0^S$) and  $\sigma^P(p)$ ($\sigma^S(p)$) denote the longitudinal (transverse)  Voigt velocity  and
longitudinal (transverse) mass operator, respectively. 
The poles of the propagators $g^P(p)$ and $g^S(p)$ give the dispersion relation of {\it P} and {\it S} waves 
in the aggregate, respectively \cite[]{Sheng95}. To determine the location of the poles,
some approximation of the mass operators has to be made. In the First-Order-Smoothing Approximation, 
correct to second order in material heterogeneity,
\citet{Weaver90} has shown that $\sigma^P(p)$ and $\sigma^S(p)$ are related to inner products
of $\Xi$ with the unit tensor $\mathbf{I}$ and the second rank tensors $\mathbf{\hat{p}}\mathbf{\hat{p}}$.
The detailed frequency dependence of the mass operators is governed by the correlation function $\eta$.
 At sufficiently low frequency, the location of the poles can be given explicitly  using the Born approximation
which consists in substituting  $p$ with $k_0\,=\,\omega/V_0$ in $\sigma(p)$ \cite[]{Weaver90}.
One obtains the following expression of the effective wave vector $k_e$ in the untextured aggregate:
\begin{eqnarray}
k_e^{P/S} \;=\; \left[\left(\frac{\omega}{{V_0^{P/S}}}\right)^2+\frac{\sigma^{P/S}(\omega/V_0^{P/S})}{\left(V_0^{P/S}\right)^2}\right]^{1/2}
\end{eqnarray}
When the Born approximation fails at high frequencies, one can still locate approximately the pole
by calculating the density of states in the random medium. This procedure is described in detail in the book of
 \citet{Sheng95}, p.85-86.
In this work, Born approximation has been used to facilitate calculations and its validity has been verified
a posteriori.  

The  effective velocities $V_e^{P/S}$ and attenuations $\alpha^{P/S}$ of the seismic waves are related to the real and imaginary parts
of the  mass operators, respectively:
\begin{eqnarray}
 \displaystyle \frac{1}{V_e^{P/S}} &=&\frac{1}{V_0^{P/S}}+\frac{1}{2\,\omega^2\,V_0^{P/S}}\,\Re \{\sigma^{P/S}(\omega/V_0^{P/S})\} \\
 \alpha^{P/S}&=&\frac{1}{2\,\omega\, V_0^{P/S}}\,\Im \{\sigma^{P/S}(\omega/V_0^{P/S})\} .   \label{born}
\end{eqnarray}
Because the material is heterogeneous at the wavelength scale, the effective phase velocity in the aggregate is
 different from the Voigt velocity.  The attenuations $\alpha^{P/S}$ 
reflect the amplitude decay of the coherent wave caused by scattering. 
In a recent review paper, \citet{Thompson02} shows the remarkable agreement between experimental
 measurements of attenuation in untextured metals and the theoretical prediction (\ref{born}).
To ease the comparison with 
seismic observations, we  define {\it P}-wave and {\it S}-wave attenuation quality factors as:
\begin{eqnarray}
Q_{P/S} =\frac{\omega}{2 \alpha^{P/S} V_0^{P/S}}
\end{eqnarray}

\section{Effective seismic properties of aggregates: study of 6 iron models from the literature }
\subsection{Anisotropy and Voigt speed of single crystals}
\begin{figure}
\center
\includegraphics[angle=0,width=15.0cm]{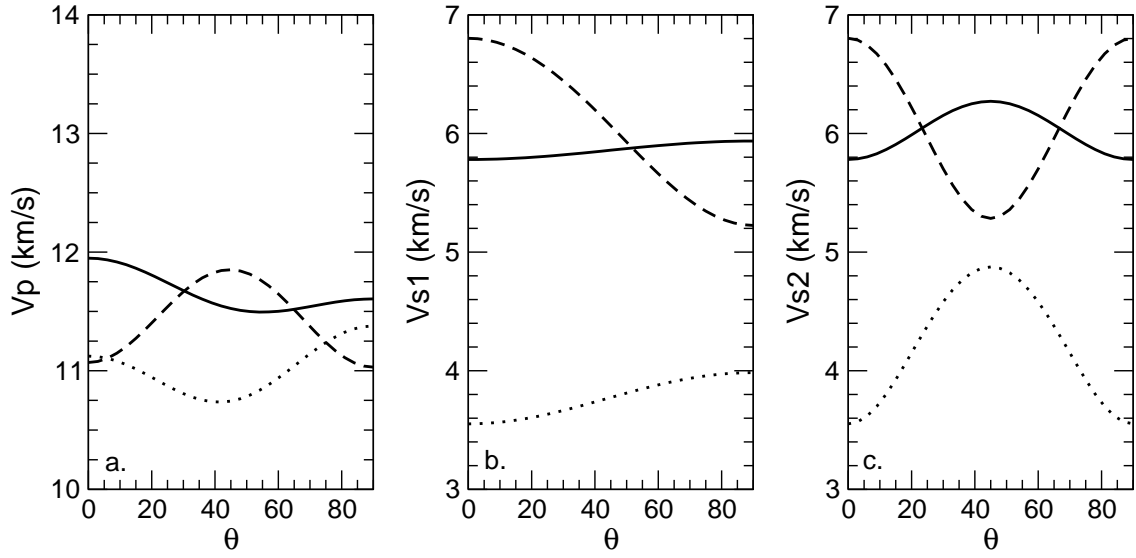}
\caption{{\it P}-wave velocity (a) and {\it S}-wave velocities (b and c) in a single crystal of iron as a
 function of propagation direction with respect to symmetry axis, for 3 iron models:
 \citet{Laio00} (solid line), \citet{Mao98} (dashed line) and \citet{Voca07} (dotted line).
$\theta$ is the angle between the direction of propagation and the symmetry axis of the crystal.}
\label{comp_vit}
\end{figure}

 Table \ref{const_elas} gives elastic properties of 
six iron crystals proposed for the inner core, obtained from either laboratory experiments
or theoretical calculations.
 These crystals present very different anisotropic characteristics as 
 revealed by Table \ref{const_elas}.
For example, the sign and amplitude of the anisotropic parameters for hexagonal iron defined in  Appendix A
vary significantly from one crystal to an other (Table \ref{const_elas}).
This is illustrated in Figure \ref{comp_vit} where the {\it P} and {\it S} wave velocities
of three hexagonal iron crystals are plotted as a function of the direction of propagation
 measured from the symmetry axis. 
 In Figure \ref{comp_vit}a, the longitudinal sound velocity
  deduced from x-ray diffraction experiments \cite[]{Mao98} has a maximum at 45$^{\circ}$
from the symmetry axis. This result has not been supported by either first principle calculations
 or other experimental studies.
The theoretical investigations themselves have led to contradictory conclusions:
the symmetry axis could be either fast \cite[]{Laio00} or slow \cite[]{Voca07} as shown in Figure \ref{comp_vit}.
Recently, \citet{Dubro07} have proposed that a body-centered-cubic ({\it bcc}) phase could be stable at inner core conditions.
Recent cubic iron models computed by \cite[]{Belono07} and \citet{Voca07} at inner core temperature,
 present different anisotropic characteristics, described by the parameter $\nu$ in Table \ref{const_elas} (Appendix A).
All iron crystals in Table \ref{const_elas} have a  higher S-wave Voigt velocity
than that inferred from seismology -between 3.5~km.s$^{-1}$ and 3.67~km.s$^{-1}$ in the ak135 model  \cite[]{Kennet95} -.
Recent theoretical calculations of elastic properties of iron at inner core conditions  converge 
towards S-wave Voigt velocities between 4.0~km.s$^{-1}$ and 4.4~km.s$^{-1}$. 
\subsection{Seismic wave dispersion}
\begin{figure}
\centerline{\epsfig{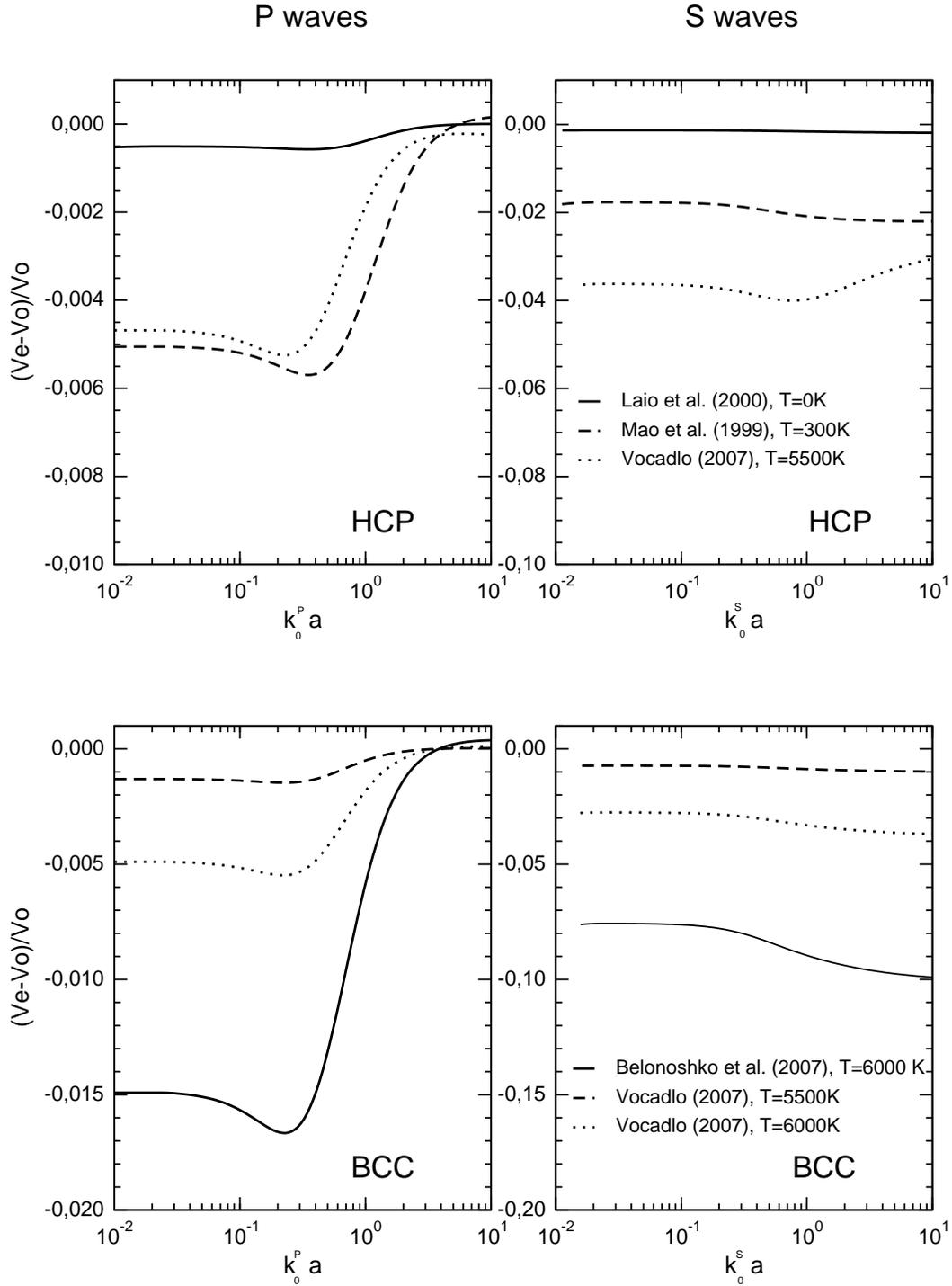}}
\caption{Normalized variations of longitudinal-wave (left) and transverse-wave (right) phase velocities in hexagonal
iron aggregates (top) and in cubic iron agregates (bottom) as a function of  adimensional
 frequency $k_0 \, a$. $k_0$ and $V_0$ are the Voigt wave vector and velocity respectively, $V_e$ is the effective velocity in the aggregate, and $a$ is the correlation length.}
\label{dispersion}
\end{figure}

We explore the seismic properties of {\it hcp} and {\it bcc} iron aggregates at the typical frequency of short period
 {\it PKIKP} waves (around 1~Hz). In what follows, an iron model will be simultaneously characterized
 by its single crystal and  aggregate properties.
  We have calculated the  {\it P} and {\it S} effective velocities for patch sizes
 ranging from 30~m to 100~km. As explained in section 2, the material heterogeneity 
 induces a shift of  the seismic wavespeeds from the Voigt velocities $V_0$. This effect is illustrated  
in Figure \ref{dispersion},  where we have plotted the normalized variation
 of phase  velocity $\delta V\,=\,(V_e-V_0)/V_0$
 as a function of adimensional frequency
$k_0 \, a$ (where $k_0$ is the Voigt wavenumber and $a$ is the correlation length) 
for {\it hcp} aggregates (top) and {\it bcc} agregates (bottom).
Over the whole frequency range, the effective velocity of both {\it P} and {\it S} waves is reduced compared to
the Voigt average. Typically, the shift is stronger by one order of magnitude for {\it S} wave than for {\it P} wave.  
For a given polarization the shift grossly increases with the degree of inhomogeneity $\epsilon^2$ (see equation \ref{voigt}), 
and  varies over one order of magnitude at least, depending on the iron models. The effective velocity is also a sensitive
function of the size of the patch. The strongest variations occur around $k^{S/P}_0 a \approx 1 $ which  
corresponds to a strong coupling between {\it P} and {\it S} modes. The ``jump''
of the velocity dispersion  curve around $k^{S/P}_0 a \approx 1 $  is absent in scalar wave propagation and 
is a remarkable feature of the elastic propagation in metals as illustrated by \citet{Stankekino}.  
At very low frequency,
our calculations verify the Hashin-Strikman bounds as they should. 
Calculations with a gaussian two-point correlation function $\eta$ yield qualitatively very similar results. 
Since gaussian correlations are usually too smooth to represent natural materials, we have preferred the more realistic exponential function. 

 Figure \ref{dispersion} illustrates the renormalization of the velocity in the heterogeneous material.
Since the effective seismic velocity of an aggregate is lower than its Voigt average velocity,
we can conclude that stable iron phases at inner core condition should have {\it P}-wave Voigt velocity
 close to the observed one, or slightly larger (around 2~\%) if the degree of inhomogeneity is large.
In the case of {\it S} waves, the difference between the Voigt and the seismic 
velocity has to be positive, and can be  large (up to 10~\% for the model of \citet{Belono07}).
 Our calculations show that the  Voigt velocity is not the appropriate
parameter to discuss the compatibility of iron models obtained from mineral physics with seismic observations.

\subsection{Seismic wave attenuation}
\begin{figure}
\center
\includegraphics[width=13.5cm]{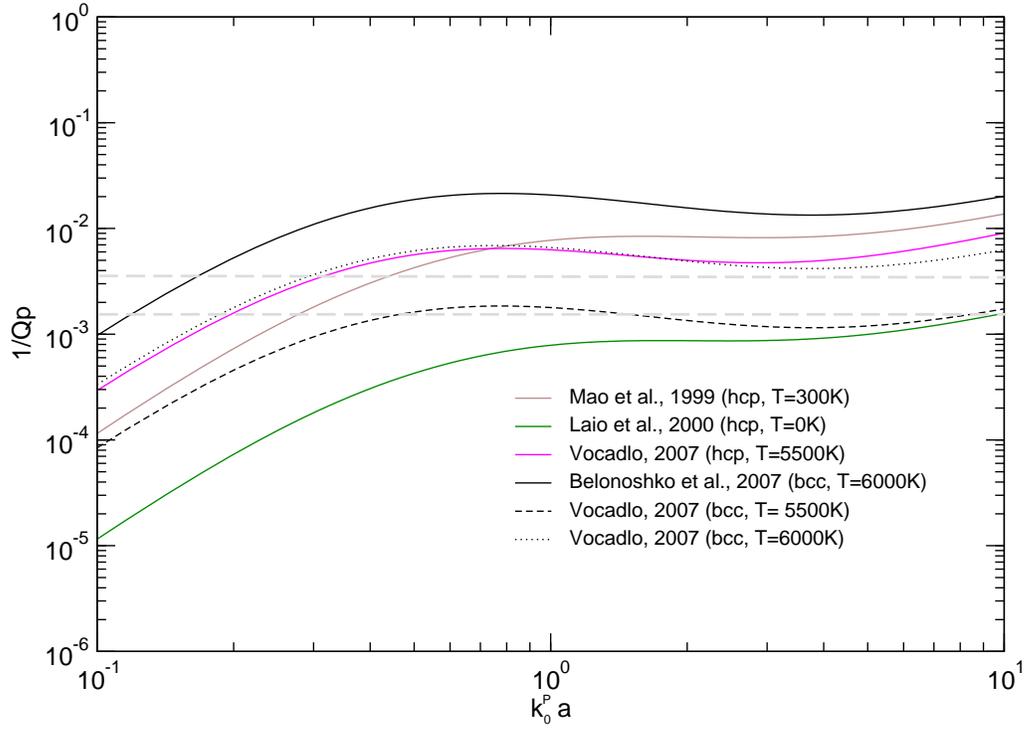}
\caption{{\it P}-wave attenuation ($1/Q_P$) as a function of  adimensional
 frequency $k_0 \, a$. Amplitude attenuation is computed for three hexagonal
 iron models and three cubic iron models proposed for the inner core.
 The  grey dashed lines correspond to $Q_P=300$ and $Q_P=600$.} 
\label{comp_att}
\end{figure}

We now examine the attenuation properties of six iron aggregates whose properties are summarized in  Table~\ref{const_elas}.
In Figure~\ref{comp_att} we show the {\it P}-wave attenuation ($1/Q_P$) as a function of adimensional frequency $k^{P}_0 a $.
Variations of $1/Q_P$ with respect to adimensional frequency depend on whether 
$\omega$ or $a$ is varied. In our case, we work at constant central frequency $\omega = 2 \pi$ and let 
$a$ range from 300~m to 90 km. 
A prominent feature is the strong frequency dependence of {\it P}-wave attenuation. Typically, the attenuation
increases by about two orders of magnitude for {\it P} waves as $k_0 a$ varies from 0.1 to 10.

Scattering attenuation is very sensitive to the anisotropic
 characteristics of iron crystals. For example, the iron
 models of \citet{Mao98} and 
\citet{Laio00} have nearly equal Voigt velocities for both {\it P} and {\it S} waves (see Table \ref{const_elas}).
 However, the elastic anisotropy of the crystals is quite different as illustrated in Figure \ref{comp_vit}.
  As a consequence, we find one order of magnitude
difference for {\it P}-wave attenuation in the aggregates, at fixed patch size.
At high frequency, we observe that longitudinal attenuation increases with the degree of
inhomogeneity $\epsilon_P^2$. All the attenuation curves
show a smooth bump around $k_0 a = 1$, which corresponds to a frequency range where the
coupling between {\it P} and {\it S} waves is maximum. Because attenuation and velocity
are related by Kramers-Kr\"onig relations, the ``jump'' of the velocity curve and the ``bump''
of the attenuation curve coincide.
 Considering these six iron models proposed for the inner core, we find as
much as two orders of magnitude difference in  attenuation for {\it P} waves at fixed patch size.
Elastic wave attenuation is therefore an important quantity to characterize
the properties of iron in the inner core. 

In Figure \ref{comp_att}, the grey lines give some typical values for {\it P}-wave
attenuation at 1~Hz at the top of the inner core, i.e. $300 < Q_P < 600$  as proposed by \citet[]{Yu06b}.
Several iron models can explain the observed $Q_P$ by scattering attenuation only, but with different textures. 
For example,  the \citet{Belono07} model
yields typical size of patches of about a few hundred meters, whereas
the {\it bcc} iron model at 5500~K by \citet{Voca07} implies patches around ten kilometers.
In the case of  \citet{Laio00}, it seems difficult to explain {\it P}-wave attenuation
 in the uppermost inner core by scattering only.
The study of these six iron models reveals a clear trade-off between elastic
properties and typical size of patches (scattering objects). 
In what follows, this problem will be further examined in order to
simultaneously constrain crystal anisotropy and the texture of aggregates
in the uppermost inner core.

\section{Constraints on elastic contants of iron in the inner core}
In this section, we consider the following inverse problem:
find iron models that match the seismic measurements of {\it P} and {\it S} wave velocities,
and {\it P} wave attenuation at 1 Hz in the uppermost inner core. An iron model is defined
by the elastic tensor of a single crystal and the texture of the aggregate.
 Because the stable phase of iron at inner core conditions is still uncertain, we
 propose to examine two kinds of crystal symmetry: hexagonal and cubic.
 In our simple model, the texture is characterized by the correlation length $a$, equal to half
 the average patch size. We let $a$ vary over more than two orders of magnitude: 50~m $<a<$ 10000~m.
 
 The observation of strong {\it PKiKP} coda suggests that scattering attenuation is the dominant
 mechanism for attenuation in the inner core. This interpretation is also supported
 by the study of \citet{cormli02} and \citet{Cormier07}. A commonly accepted value for
 $Q_P$ at the top of the inner core is about 300 \cite[]{Yu06b}. We impose that an
 acceptable iron model should have a scattering $Q_P$ within the range 300-600. This choice implies
 that more than half the observed attenuation is explained by scattering and leaves
 some room for other attenuation mechanisms.  Since {\it S}-wave attenuation in the inner core around 1 Hz is still poorly
 known, we do not impose  constraints on $Q_S$. The range of acceptable seismic velocities corresponds 
 to the values given by the ak135 model for the whole inner core.
\subsection{Hexagonal iron}
\begin{figure}
\centerline{\epsfig{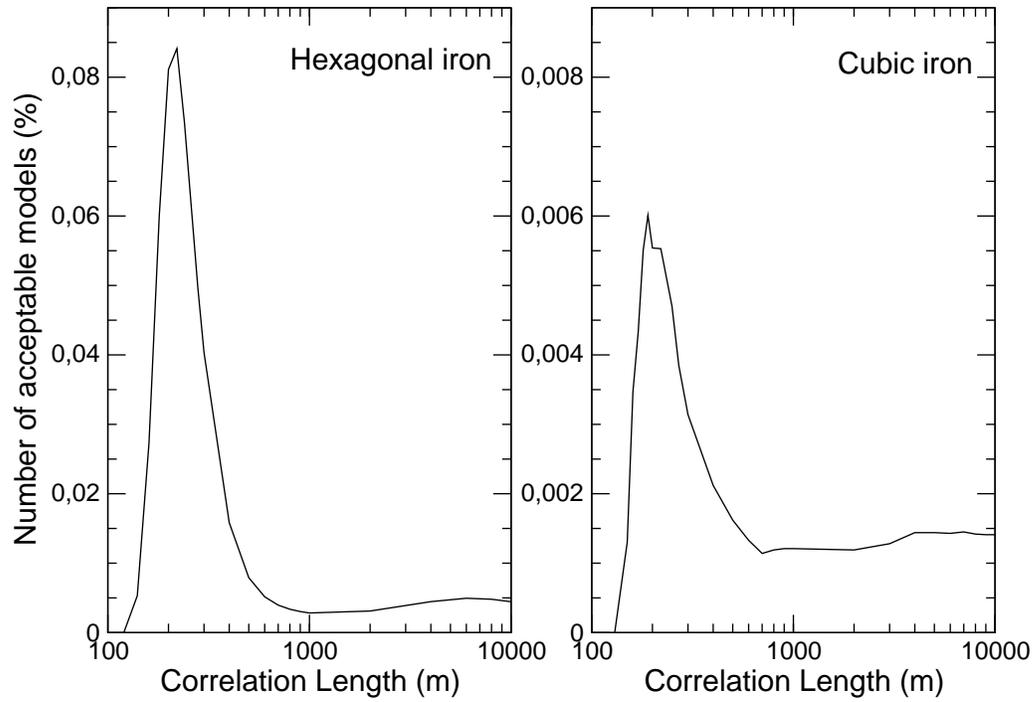}}
\caption{Number of acceptable iron crystals as a function of the correlation length: hexagonal symmetry (left) and
cubic symmetry (right). The corresponding aggregates verify
the seismic observations at 1 Hz: P-wave and S-wave velocities are in the range of the  ak 135 model of the whole inner core, and attenuation quality factor $Q_P$ lies between  300 and 600.}
\label{nhex}
\end{figure}

To find acceptable iron models we perform a systematic grid search
in the $C_{IJ}$ space. Only positive definite elastic tensors have been retained.
For each correlation length, we test 1.8$\times 10^8$ physical iron models with 
anisotropy parameters $(\varepsilon,\delta,\gamma)$ smaller than 50\%.
 In  Figure \ref{nhex} (left), we represent the percentage of acceptable hexagonal models as a function
 of the correlation length. The curve shows a sharp maximum  
at $a\simeq220$~m, which yields an  average size of the patches around 450~m. The correlation
length of the aggregate cannot be smaller than 140~m but can be as large as 10000~m. It is to be noted
that the most probable patch  size is about 20 times smaller than the central wavelength of {\it P} waves.
  A careful analysis reveals that the bounds on the
seismic velocities determine the overall number of acceptable
models, whereas  the bounds on the attenuation fix the position of the maximum. 

\begin{figure}
\center
\includegraphics[width=15.0cm]{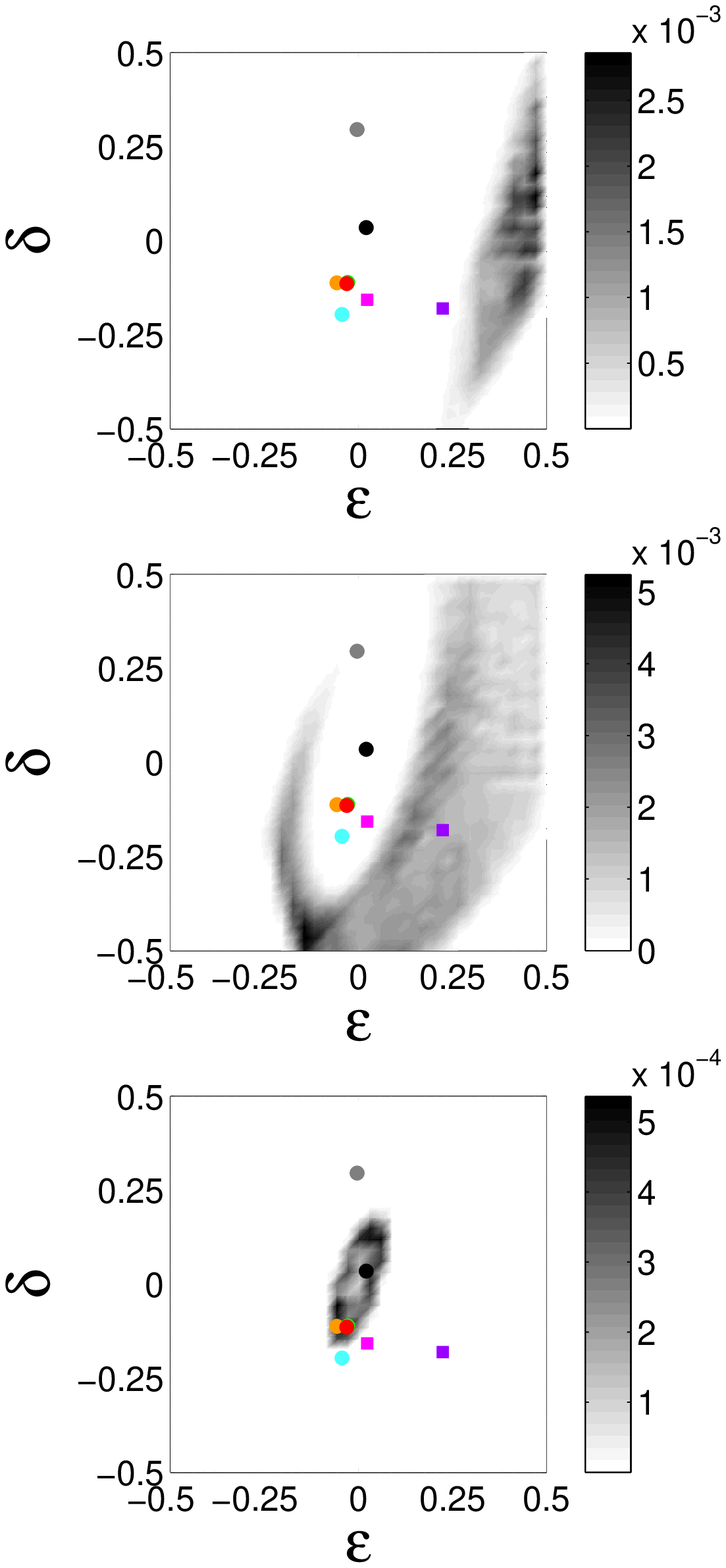}
\caption{Density of acceptable hexagonal crystals as a function  of the anisotropic parameters $\varepsilon$ and $\delta$, for $a = 140$~m (top), $a = 220$~m
(middle) and $a = 1000$~m (bottom). For comparison, color dots representing
eight {\it hcp} iron models are plotted:
\citet{Stix95} (red), \citet{Soder96} (black), \citet{Mao98} (brown), \cite{Stein99} (cyan),
 \citet{Laio00} (green), \citet{Stein01} (purple), \citet{Voca03a} (orange), \citet{Voca07} (magenta).
 Only two models have been computed at inner core temperature (squares).}
\label{distrib_hexa}
\end{figure}

\begin{figure}
\centerline{\epsfig{file=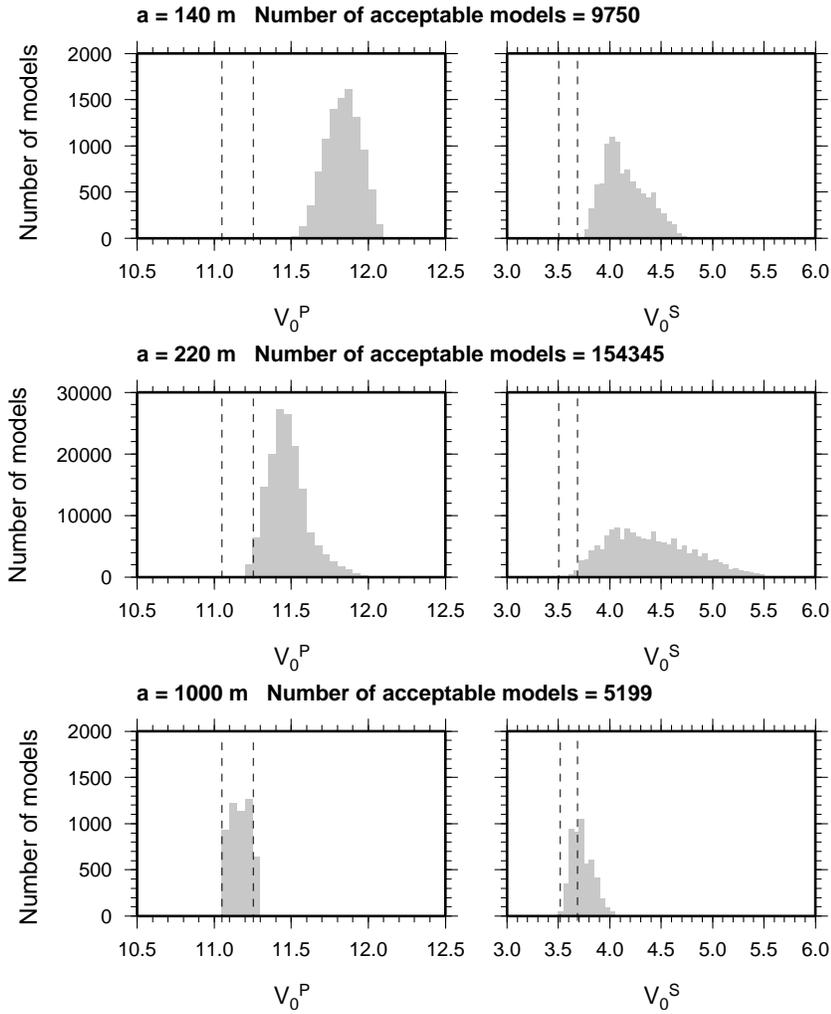,width=0.8\linewidth,bbllx=0,bblly=0,bburx=385,bbury=466,clip=}}
\caption{Distribution of Voigt velocities of acceptable  {\it hcp} iron crystals 
 for different correlation lengths: $a=140$~m, $a=220$~m, $a=1000$~m (from top to bottom). Vertical lines delimit the range of seismic velocities in the inner core.}
\label{distrib_voigt}
\end{figure}

For three correlation lengths $a=$140~m, 220~m, 1000~m,  
we examine in Figure \ref{distrib_hexa} the  distribution of acceptable models in the ($\varepsilon$, $\delta$) parameter space.
We focus on the two parameters controlling the  {\it P}-wave crystal anisotropy, in order 
 to facilitate comparisons  with results from mineral physics. 
 Figure \ref{distrib_hexa} is a 2-D
 histogram of the number of acceptable models.
For each correlation length, there is a clear trade-off between $\varepsilon$ and $\delta$ because
 only three independent observations serve to constrain 5 independent parameters. 
  To avoid artefacts due to uneven sampling of
 the model space, we have taken care of normalizing the results by the total number of tested models
 per class. In fact the systematic exploration of the $C_{IJ}$ space yields relatively homogeneous coverage of the
 ($\varepsilon$, $\delta$) space. In Figure \ref{distrib_voigt}, we show  the a-posteriori histograms of acceptable Voigt 
  velocities for the three correlation lengths. 

 At small correlation length ($a=140$~m), the aggregate is most likely composed of
 hexagonal iron crystals with a slow symmetry axis ($20\%<\varepsilon<50\%$),
 and possibly high Voigt velocities (11.6~km.s$^{-1}$$<V_0^P<$12.1~km.s$^{-1}$, 3.8~km.s$^{-1}$$<V_0^S<$4.7~km.s$^{-1}$).
 There is no clear constraint on $\delta$.  We remark that  Voigt velocities of iron crystals
 can never be smaller than seismic velocities, but high Voigt velocities
  are not necessarily incompatible with seismic observations.  

At large correlation length  ($a  > 500$~m), the ($\varepsilon$ , $\delta$) distribution 
 has the shape of an ellipse centered around $(0,0)$, with positive correlation
 between $\varepsilon$ and $\delta$ in the range  $-10\% < \varepsilon < 10 \%$, $ -20\%<\delta<25\% $.
 The crystal Voigt velocities must be close to seismic velocities.
Careful analysis shows that there are slightly
more {\it hcp} crystals with a slow symmetry axis, than with a fast symmetry axis.
At large correlation length, none of the 8 hexagonal iron models considered in this study has  
 both anisotropic parameters and Voigt velocities compatible with seismic observations.

For $a\sim220$~m, we obtain a maximum of acceptable iron aggregates.
Corresponding iron crystals have a  complex distribution in 
$\varepsilon$ and $\delta$. $\delta$ is not constrained,
and $\varepsilon$ varies between $ -20\%$ and  $ +50\%$. Interestingly, the elliptic
domain which defined acceptable models at large correlation length is forbidden
when $a$ equals the most probable patch size.
 Like for small correlation lengths, we obtain a larger number of hexagonal iron crystals with
 a slow symmetry axis for P waves.
The distributions of acceptable  P-wave Voigt velocity
 (11.3~km.s$^{-1}$$<V_0^P<$12.0~km.s$^{-1}$) and  S-wave Voigt velocity 
(3.5~km.s$^{-1}$$<V_0^S<$5.5~km.s$^{-1}$) are very broad, with an average
clearly higher than the seismic velocities.
The iron crystal proposed by \citet{Stein01} is the only model which presents 
both anisotropic parameters ($\varepsilon,\delta$) and Voigt velocities that 
are compatible with seismic observations. However, this iron model is questionable  
because of the too large value of the axial ratio of {\it hcp } iron predicted by the calculations \citep{Ganna05}.

 \subsection{Cubic iron}
We perform a grid search in the $(C_{11},C_{12},C_{44}) /\rho$ parameter space, where $\rho$
is the model iron density. The normalization by $\rho$  facilitates the comparison between models
with different densities. In any case, the scattering properties of an aggregate depend only on $C_{IJ}/\rho$. 

For a given texture, we test 8.8$\times10^6$ cubic iron crystals with a homogeneous distribution
in the range: $0.1 < C_{11}/\rho < 0.2$, $0.06 < C_{12}/\rho < 0.12$, $0 < C_{44}/\rho < 0.04$, where
the $C_{IJ}$ and $\rho$ units are GPa and kg.m$^{-3}$, respectively.
In  Figure \ref{nhex} (right), we represent the percentage of acceptable cubic models as a function
 of the correlation length. The curve shows a  maximum  
at $a\simeq200$~m, which yields an  average size of the patches around 400~m. 
Therefore the most probable size of  patch is a robust feature, independent of the symmetry
of the crystal. Like in the hexagonal case, the correlation
length of the aggregate cannot be smaller than 150~m but can be as large as 10000~m. 
 The percentage of accepted  models is by one order of magnitude smaller
in the cubic case than in the hexagonal case. 

\begin{figure}
\centerline{\epsfig{file=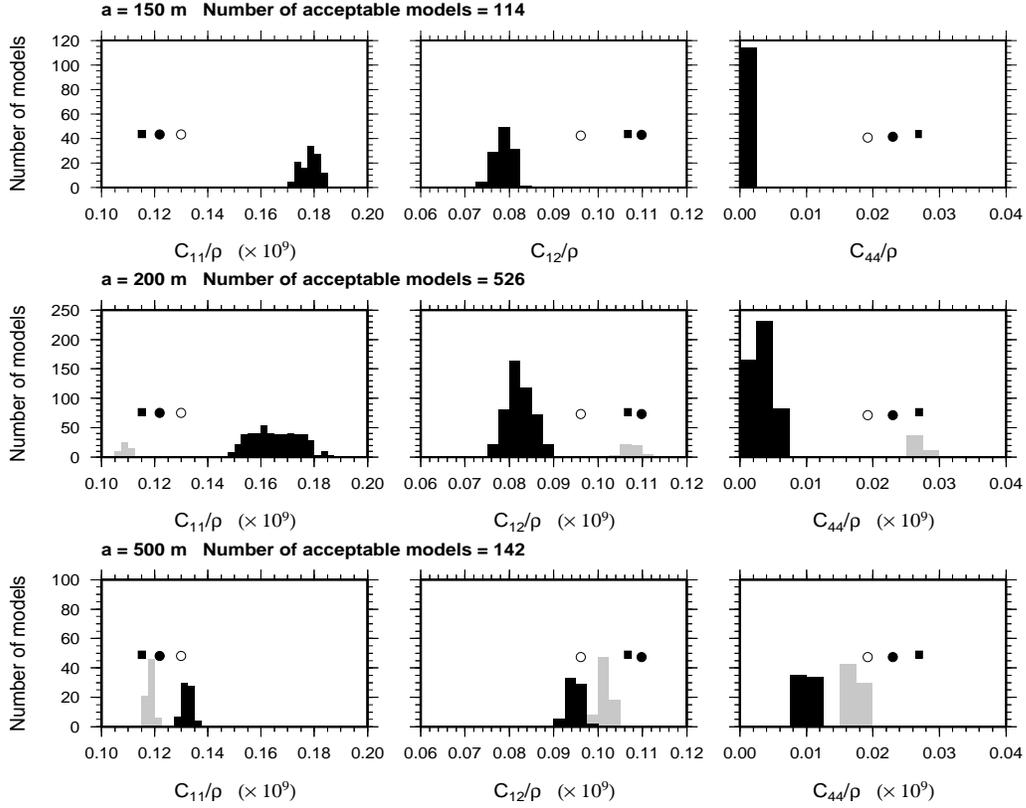,width=1.0\linewidth,bbllx=-10,bblly=-10,bburx=650,bbury=530,clip=}}
\caption{Distribution of acceptable elastic constants $C_{IJ}/\rho$ of cubic iron for different correlation lengths, indicated on top of each plot.
At large correlation length, we observe two distinct families of cubic models (shown in grey and black).
Circles correspond to two {\it bcc} iron models proposed by \citet{Voca07} for a density $\rho =$~13155~kg.m$^{-3}$ and a temperature T$=$5500~K (black), and for $\rho =$~13842~kg.m$^{-3}$ and T$=$6000~K (white). The square corresponds to the {\it bcc} iron model proposed by \citet{Belono07} for a density $\rho=$~13580~kg.m$^{-3}$ and a temperature T$=$6000~K.}
\label{hist_cub}
\end{figure}

For three correlation lengths (150~m, 200~m and 500~m), 
we examine, in Figure \ref{hist_cub}, the distribution
 of   elastic constants of  acceptable cubic crystals. 
 At large correlation length, we observe two distinct families  shown in grey and black
 in Figure  \ref{hist_cub} (middle-bottom), whereas a single family
exists at correlation length smaller than 200~m.
The black and white circles
 show the recent results of \citet{Voca07} for cubic iron at two different inner core temperatures,
and the black square is the recent {\it bcc} iron model by \citet{Belono07}. 
 This last  model is completely compatible with seismic observations and corresponds to the ``grey'' family of cubic models. 

\begin{figure}
\centerline{\epsfig{file=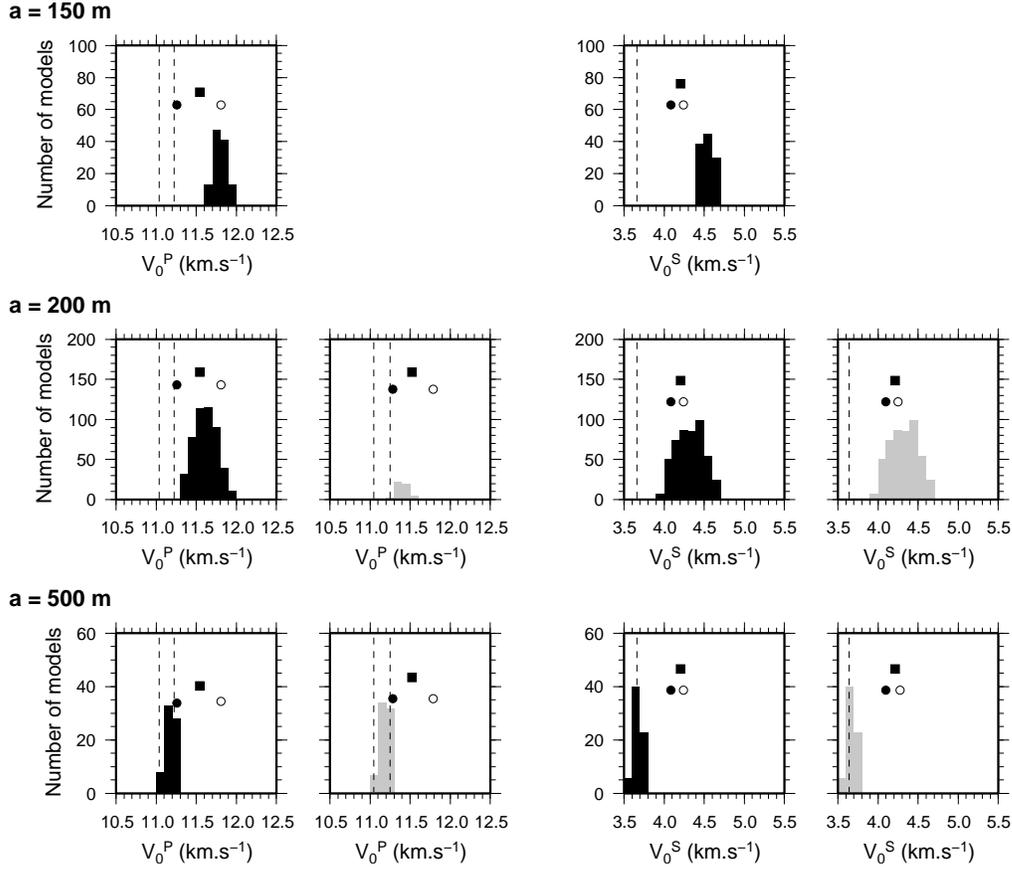,angle=90,width=1.0\linewidth,bbllx=-10,bblly=-10,bburx=466,bbury=545,clip=}}
\caption{Distribution of  Voigt velocities for the two families of acceptable cubic models of iron shown in Figure 8. 
The correlation length is indicated on top of each plot: $a= $150, 200, 500~m. Circles correspond to the two {\it bcc} iron models proposed by \citet{Voca07} and the square is the {\it bcc} model proposed by \citet{Belono07} (same symbols as in Figure 8). Vertical lines delimit the range of seismic velocities in the inner core. }
\label{qs_cub}
\end{figure}

In Figure \ref{qs_cub}, we show the a posteriori distributions of Voigt velocities for  three correlation lengths.
The main features are very similar to the hexagonal case but the
bounds are slightly sharper. For the most probable patch size, the average  {\it S}-wave
Voigt velocity is around 4.4 km.s$^{-1}$, in agreement with recent results by \citet{Voca07}
and \citet{Belono07}.  
Our study shows that some cubic iron crystal with S-wave Voigt velocity around 4.0~km.s$^{-1}$
 can explain seismological data without invoking the presence of melt, as for example the iron model by
 \citet{Belono07}. On the other hand the {\it bcc} iron model at $5500K$ by \citet{Voca07}, 
although in reasonably good agreement with the 
results of our inversion, is not
 anisotropic enough to induce a strong renormalization of the seismic velocities in the aggregate (only 1~\%
 as shown in Figure \ref{dispersion}). For such an iron model, it is necessary 
 to invoke the presence of some melt to explain the seismic velocities.
%
\section{Discussion and conclusion} 
\subsection{Patch size}
 (1) From mineral physics and geodynamics. Estimates of the size of iron crystals in the Earth's
 inner core vary from the entire volume,
 some 1200~km, to about 5~mm. \citet{Stix95} have proposed
that a single crystal of iron with anisotropy around 4~\% could explain seismic traveltimes.
The smaller estimates stem from geodynamical considerations. Near the melting temperature
of iron, a grain size of about 5~mm is necessary to make the viscosity smaller than $10^{16}$~Pa.s,
as proposed by \citet{Buf97}. Assuming dynamic
 recrystallisation, \citet{yoshida96} obtained a typical crystal size around 5~m.
\citet{Bergman97,Bergman98} suggested that observed attenuation and velocity anisotropy of the inner core may result
from radially elongated columnar grains and estimated
the columnar grain width around 200~m.  This last study is in good  agreement with our most probable
patch size.

(2) From seismology. With scattering and a fabric interpretation of seismic attenuation, 
\citet{cormli02} have proposed average scale length of
isotropic heterogeneities around 10~km, and {\it P}-wave velocity perturbations around 8.4\% 
 from the centre of the inner core to about 1000~km radius. From the energy envelopes  of {\it PKiKP} coda waves,
\citet{VidaleNat} have obtained a scale length of about 2~km, and elastic moduli fluctuations of
 1.2\% in the uppermost 300~km of the inner core. These results may reflect a trade-off 
 between velocity perturbations and scale length. However, the most serious criticism is the lack
 of physical connection between the hypothesized isotropic seismic heterogeneities and the crystalline nature of
 the inner core. Our most probable correlation length (around 200m)  is smaller than previous estimates by \citet{cormli02} and \citet{VidaleNat},
  but is close to the lower limit proposed by \citet{Cormier98} (about 15 \% {\it P}-wave perturbations  and
 scale length in the range 0.5-2~km). 
  
\subsection{On the presence of melt}
The vast majority of proposed iron models for the inner core displays
much higher Voigt average velocities than what is inferred from seismic
traveltimes. This discrepancy has sometimes been used as an argument to support
the presence of liquid in the inner core \cite[]{Singh00,Voca07}. In our study, we have
shown that high  Voigt velocities, typically of the order of what is found in the most
recent studies \cite[]{Belono07,Voca07}, do not necessarily imply the presence of melt. The reason is
that the effective velocity in an aggregate can be as much as 15 \% smaller
 (according to the anisotropic characteristics of iron crystal) than the
average velocity in a single crystal. Two additional points have to be noted: (1) Iron
crystals with Voigt velocities smaller than seismic wavespeeds have to be rejected,
independent of our assumption for the value of $Q$ (2) Iron models
with Voigt velocities larger than 5.5km.s$^{-1}$ can also be  clearly rejected, unless they
contain a huge amount of melt, typically more than 20 \% according to Figure~2 in \citet{Singh00}.

\subsection{Cubic vs Hexagonal}
Most theoretical and experimental studies show that hexagonal metals at ambient temperature 
have a fast symmetry axis \cite[]{Antona06}. But the effect of temperature on the elastic constants
is still debated \cite[]{Laio00}. We found that {\it hcp} iron models with a slow symmetry axis are more probable
 in the inner core. If further experimental or theoretical
investigations definitely show that {\it hcp} iron at inner core conditions
 has a fast symmetry axis, we may conclude on the basis of our uppermost 
inner core model that the stable iron phase
 is unlikely to be hexagonal. Recent cubic models from the literature are in very good agreement with the results of our 
inversion.  Therefore, our study   supports a stable {\it bcc} iron phase in the inner core,
 as proposed by \citet{Dubro07}.
  
 Although our study provides some indications on the possible elastic constants of iron, some large uncertainties remain. They
 are clearly caused by a lack of seismological constraints.
  In future studies, the modeling of the following three observations could significantly reduce the uncertainties 
 on the properties of  iron in the inner core: 
 (1) energy envelope of PKiKP coda; (2) direct measurements of $Q_S$ at 1~Hz; (3) velocity and attenuation
 anisotropy in the bulk of the inner core.  As a large number of records of {\it PKiKP} coda are available, for which
 multiple scattering may be relevant, we plan to focus our efforts on this aspect in future works. 
 The modelling of  frequency dependent inner core attenuation 
combined with coda analysis could unravel the
relative contributions of viscoelasticity and scattering attenuation.
\bibliography{biblio_tot}

\begin{appendix}
\section{Anisotropic parameters for hexagonal and cubic crystals}
\subsection{Hexagonal symmetry}
In a transversely isotropic medium (hexagonal symmetry),
the symmetry axis defines a particular direction
 which is usually chosen along the $\mathbf{\hat{x}_3}$-axis for convenience.
To describe anisotropic properties of hexagonal crystals, \citet{Mensch97} have defined
three anisotropic parameters:
\begin{eqnarray}
\varepsilon &=& \frac{(C_{11}-C_{33})}{2C_{33}} \nonumber \\
\delta &=& \frac{(C_{13}-C_{33}+2C_{44})}{C_{33}}\\
\gamma &=& \frac{(C_{66}-C_{44})}{2C_{44}} \nonumber
\label{parameter}
\end{eqnarray}
The parameter $\varepsilon$ represents {\it P}-wave anisotropy $i.e.$ the difference
between the phase velocities perpendicular and parallel to the symmetry axis.
A negative  value of $\varepsilon$ corresponds to a fast symmetry axis for {\it P} waves (Table \ref{const_elas}).
 The parameter $\gamma$  represents the anisotropy of {\it S} waves in a similar way.
 The parameter $\delta$ controls {\it P} wave propagation
 at intermediate angles from the symmetry axis.
In order to fully prescribe the elastic properties of hexagonal crystals, two additional
parameters have to be introduced.  We have chosen the  {\it P} and {\it S}  Voigt average velocities
because they are routinely compared with the seismic measurements.
\subsection{Cubic symmetry}
The stiffness tensor of a cubic crystal is described by three independant
constants $C_{11}$, $C_{12}$ and $C_{44}$. The invariant anisotropy 
factor for cubic-class crystal is defined as:
\begin{eqnarray}
\nu = C_{11}-C_{12}-2C_{44}
\end{eqnarray}
The elastic properties of cubic crystal are fully described by the parameter $\nu$ and the {\it P} and {\it S}  Voigt average velocities.
\end{appendix}
%

%
\newpage
%
%
\pagestyle{empty}
\begin{landscape}
\begin{table}
\begin{center}
\begin{tabular}{|l||c|ccccc||cccccc||cc|} \hline
Hexagonal Iron Models & $\rho$ & $C_{11}$&$C_{12}$&$C_{44}$&$C_{33}$&$C_{13}$&$V_0^P$&$V_0^S$&$\nu$&$\varepsilon$&$\delta$&$\gamma$&$\epsilon_P$&$\epsilon_S$\\
  & ($kg.m^{-3}$) & (GPa) & (GPa) & (GPa) & (GPa) & (GPa) & (km.s$^{-1}$) & (km.s$^{-1}$) &(GPa)&(\%) & (\%) & (\%) & (\%) & (\%)\\
\hline
\hline
\multicolumn{15}{|l|}{HCP Iron} \cr
\hline
\citet{Mao98}& 12600 & 1533 & 367 & 583 & 1544 & 835 & 11.47 & 5.92 & & -0.36 & +29.60 & -20.50 & 2.46 & 8.71\\
\citet{Laio00}& 12885 & 1697 & 809 & 444 & 1799 & 757 & 11.45 & 5.90 & & -2.84 & -11.12 & +2.73 & 0.80 & 1.75\\
\citet{Voca07}& 13155 & 1311 & 159 & 1642 & 1642 & 1074 & 11.14 & 4.04 & & +2.84 & -15.22 & +15.72 & 2.22 & 5.17 \\
\hline
\hline
\multicolumn{15}{|l|}{BCC Iron}\cr
\hline
\citet{Belono07}& 13850 & 1561 & 1448 & 365 & & & 11.54 & 4.20 & -617 & & & & 2.79 & 17.53 \\
\citet{Voca07}& 13155 & 1603 & 1258 & 256 & & & 11.29 & 4.11 & -167 & & & & 0.86 & 4.91 \\
\citet{Voca07}& 13842 & 1795 & 1519 & 323 & & & 11.83 & 4.24 & -370 & & & & 1.60 & 9.73 \\
\hline
\end{tabular}
\end{center}
\caption{Elastic properties of hexagonal ({\it hcp}) and cubic ({\it bcc}) iron crystals and associated untextured aggregate:
 density ($\rho$), elastic constants ($C_{IJ}$), hexagonal anisotropic parameters  ($\varepsilon$,$\delta$,$\gamma$), 
cubic anisotropic parameter ($\nu$), {\it P}-wave and {\it S}-wave Voigt velocities ($V_0^P, V_0^S$), degrees of
 inhomogeneity ($\epsilon_P^2$,$\epsilon_S^2$). The iron models have been computed at inner core temperature
 except \citet{Mao98} and \citet{Laio00}.}
\label{const_elas}
\end{table}
\end{landscape}
\newpage
\newpage
%

\end{document}